\def\be{\begin{equation}}
\def\ee{\end{equation}}
\def\kt{k_{\perp}}
\def\pt{p_{\perp}}
\def\<{\langle}
\def\>{\rangle}

\newcommand{\bea}{\begin{eqnarray}}
\newcommand{\eea}{\end{eqnarray}}
\newcommand{\nn}{\nonumber\\}

\documentclass{PoS}

\title{Studies of TMDs with CLAS}

\ShortTitle{Studies of TMDs with CLAS}

\author{\speaker{M. Aghasyan}\\
        LNF, INFN, Via E. fermi 40, Frascati (RM) 00044, Italy\\
        E-mail: \email{aghasyan@lnf.infn.it}}

\author{H. Avakian\\
        JLab, 12000 Jeferson Ave, Newport News, VA 23606, USA\\}

\abstract{Studies of single and double-spin asymmetries in pion electro-production in semi-inclusive deep-inelastic scattering of 5.8 GeV polarized electrons from unpolarized and longitudinally polarized targets at the Thomas Jefferson National Accelerator Facility using CLAS discussed. We present a Bessel-weighting strategy to extract transverse-momentum-dependent parton distribution functions.}

\FullConference{XXI International Workshop on Deep-Inelastic Scattering and Related Subject -DIS2013,\\
		22-26 April 2013\\
		Marseilles,France}

\begin{document}
The study of the transverse spin structure of protons and neutrons is one of the central issues in hadron physics, with many dedicated experiments running (COMPASS at CERN, CLAS and Hall-A at JLab, STAR and PHENIX at RHIC), approved (JLab $12~{\rm GeV}$ upgrade, COMPASS-II) or planned (ENC/EIC Colliders). The transverse momentum dependent (TMD) partonic distributions (PDFs) and fragmentation functions (FFs) play a crucial role in the 3-dimensional imaging of nucleons. TMDs can be accessed in several types of experiments although, the main source of information is semi-inclusive deep inelastic scattering (SIDIS) of polarized leptons off polarized nucleons. 
Significant amounts of data on spin-azimuthal distributions of hadrons in semi-inclusive DIS, which provide access to TMDs, has been accumulated in recent years by 
several collaborations including HERMES, COMPASS, and Halls A, B and C at 
JLab\cite{Airapetian:2004tw,Alexakhin:2005iw,Alekseev:2010dm,Avakian:2010ae}.
The extraction of actual TMDs as a fuction of transverse momentum $\kt$ and $x$ 
from different single and double spin azimuthal
asymmetries is hindered by the absence of a reliable, model-independent procedure for flavor decompositions of the underlying TMDs. Various assumptions involved in preliminary extractions of TMDs from available data did not allow for credible estimates of systematic errors due to those assumptions, which also prevented credible projections of the 
statistics needed for an extraction of relevant TMDs. 
The rigorous basis for studies of TMDs in SIDIS is provided by
TMD factorization in QCD, which has been established in 
Refs.~\cite{Ji:2004wu,Collins:2004nx, Bacchetta:2008xw,Collins:2007ph} for 
leading twist 
single hadron production from a quark with transverse momentum $\kt$ smaller than the hard scattering scale $Q^2$ (i.e. $\kt^2 \ll Q^2$). 
In this kinematic domain, the SIDIS cross section  can be expressed in terms of 
structure functions that encode the strong-interaction dynamics of
the hadronic subprocess $\gamma^* + p\to h + X$  
\cite{Kotzinian:1994dv,Mulders:1995dh,Levelt:1994np,Bacchetta:2006tn,Lu:2012gu}, which are  convolutions of  transverse momentum dependent distribution and fragmentation
functions.

In the recent paper by Boer, Gamberg, Musch and Prokudin (BGMP)~\cite{Boer:2011xd} a new technique has been proposed, that allows a model-independent extraction of Fourier transforms of TMD distributions from observed azimuthal moments in SIDIS with polarized and unpolarized targets.  
A fully differential Monte Carlo (MC)  generator has been developed \cite{Aghasyan:2013qqa,Aghasyan:2013kz} to test the procedure for extraction of TMDs from SIDIS in a model independent way, based on the BGMP formalism.

Such a  Monte Carlo generator is a crucial component in testing different procedures for flavor decomposition of TMDs. The Monte Carlo generator we used has been 
developed to study partonic intrinsic motion 
within the framework generalized parton model described in Ref.~\cite{Anselmino:2005nn}. 
In SIDIS,  the theoretical formalism is described in a series of papers~\cite{Anselmino:2005nn,Boglione:2011wm} using tree level factorization~\cite{Mulders:1995dh} where  the standard momentum convolution integral relates the quark
intrinsic momentum to the  transverse momentum of the produced hadron
$P_{h,T}$.
Although based on simple assumptions, this model can be
seen as a good approximation to understand some physical QCD features
and kinematical constraints  \cite{Boglione:2011wm}. Adopting the kinematic relations described in \cite{Anselmino:2005nn}, we keep the freedom of changing the distribution and fragmentation functions to check the sensitivity of the extraction procedure. 

\begin{figure}[t]
\centerline{\includegraphics[height=1.5in]{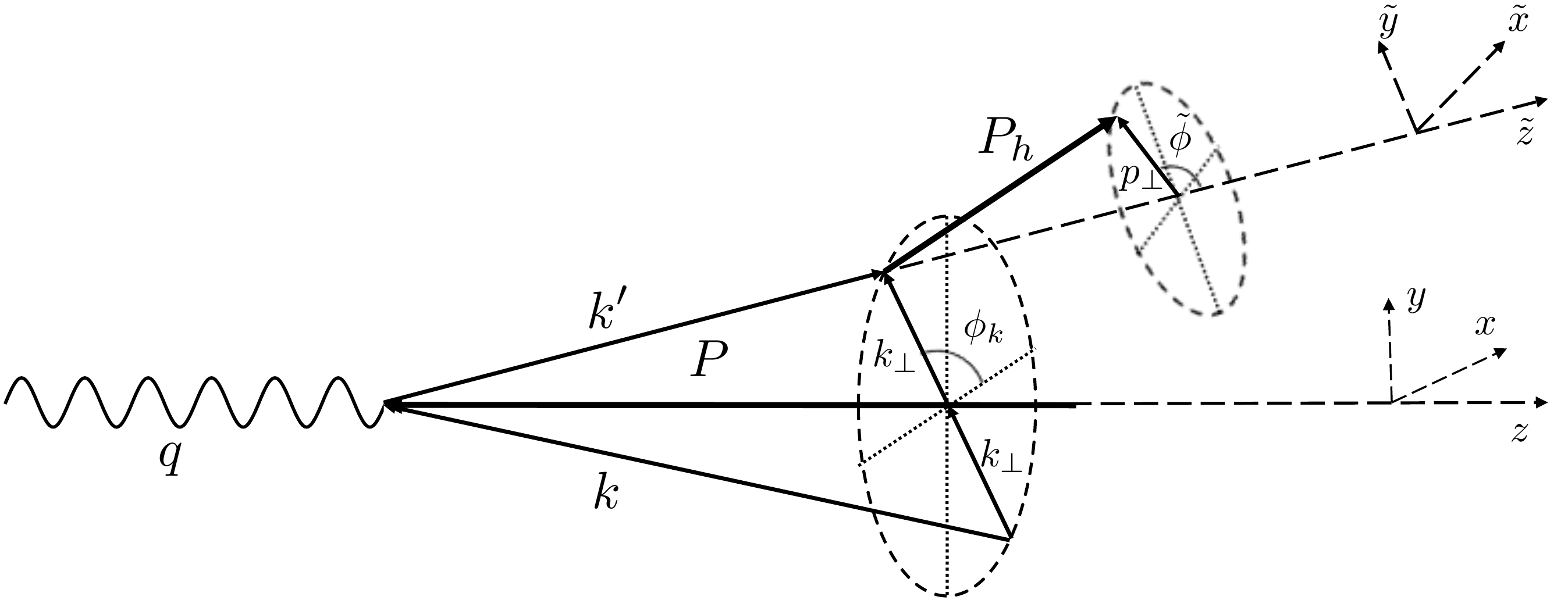}}
\caption{Kinematics of the SIDIS process. Here $q$ is the virtual photon, $k$ and $k'$ are the initial and final quarks, $k_{\perp}$ is the quark transverse component. $P_h$ is the final hadron with a $p_{\perp}$ component, transverse with respect to the fragmenting quark $k'$ direction. }
\label{kinem}
\end{figure}

We discuss the process
\be
{\ell}(l) + N(P)\rightarrow \ell(l') + h(P_{h}) + X, 
\ee
where $\ell$ is the lepton, $N$ is the proton target and $h$ is the observed hadron (four-momenta given in parentheses). 
The virtual photon momentum $q$ is along the $z$ direction and the proton momentum $P$ is in the opposite direction, as presented in Fig~\ref{kinem}. The detected hadron has momenta $P_h$. In the parton model the virtual photon scatters off a on-shell quark. The initial quark momentum $k$ and scattered quark momentum $k'$ have the same intrinsic transverse momentum component $\kt$ with respect to $z$ axis. The initial quark has only the fraction $x$ of the proton momenta in light-cone frame (see  \cite{Anselmino:2005nn} for more details). The produced hadron $P_h$ has fraction $z$ of the scattered quark momentum $k'$ in the $(\tilde x,\tilde y, \tilde z)$ frame and transverse momentum $\pt$ with respect to scattered quark $k'$.
The fully differential SIDIS cross section in MC is given by:
\bea
\frac{d \sigma}{ dx dy dzd^2 {\bf p}_{\perp}  d^2 {\bf k}_{\perp} d\phi_{l'}} &=& K(x,y)J(x,Q^2,\kt)\times \nn 
&& \hspace{ -0.5cm} \times \sum_q e_{q}^2\left[   f_{1,q}(x,k_{\perp})D_{1,q}(z,p_{\perp})  + \lambda \sqrt{1-\varepsilon^2}g_{1L,q}(x,k_{\perp})D_{1,q}(z,\pt)  \right] 
\label{FDALL}
\eea
where the summation runs over quarks flavors. The kinematic factors $K(x,y)$ and $\varepsilon$ and the Jacobian $J(x,Q^2,\kt)$ are defined in  \cite{Anselmino:2005nn}. The product of target polarization and beam helicity represented with $\lambda$ ($\lambda=\pm1$). 
The scattered lepton azimuthal angle denoted $\phi_{l'}$ and $e_q$ denotes the fractional charge of the struck quark or antiquark.

In many  phenomenological studies of semi-inclusive deep inelastic
scattering, the partonic transverse momentum dependence
 of TMDs  $ f_{1,q}(x,k_{\perp})$ and $g_{1L,q}(x,k_{\perp})$ and FFs $D_{1,q}(z,\pt)$ are  factorized from the longitudinal momentum dependence $x$ and $z$.  In more general case the dependence is assumed to be a Gaussian, with widths depending  on the fractions $x$ and $z$. 
\begin{equation}
f_{1}(x,k_{\perp})=f_1(x)\frac{1}{\pi <k^2_{\perp}(x)> _{f_1}}e^\frac{-k^2_{\perp}}{<k^2_{\perp}(x)>_{f_1}}, \qquad g_{1L}(x,k_{\perp})=\frac{g_{1L}(x)}{\pi <k^2_{\perp}(x)> _{g_1}}e^\frac{-k^2_{\perp}}{<k^2_{\perp}(x)>_{g_1}}
\label{gdf}
\end{equation}

\begin{equation}
D_{1}(z,p_{\perp})=D_1(z)\frac{1}{ <p^2_{\perp}(z)> }e^\frac{-p^2_{\perp}}{<p^2_{\perp}(z)>}
\label{gff}
\end{equation}
where $f(x)$ and $D(z)$ are given by fits from available world data and the
widths are free parameters to be extracted from the data.
In our studies we  used modified Gaussian (MG) DFs and FF from Eqs. ~\ref{gdf}-\ref{gff}, in which $x$ and $\kt$ are  
inspired by AdS/QCD \cite{deTeramond:2008ht}, with $<k^2_{\perp}(x)>=Cx(1-x)$
and $<p^2_{\perp}(z)>=Dz(1-z)$, in which constants C and D may be different for different flavors and polarization states.
 Similarly, an unfactorized DF in $z$ and $ \pt$ is also suggested by the NJL-jet model \cite{Matevosyan:2011vj}. 


We present the extraction of the double spin asymmetry $A_{LL}$, defined as the ratio of the difference and the sum of electroproduction cross sections for antiparallel, $\sigma^+$, and parallel, $\sigma^-$, configurations of lepton and nucleon spins, using the Bessel-weighting procedure described in \cite{Boer:2011xd} and applied in \cite{Lu:2012ez}.
Within this approach, one can extract the Fourier transform of the double spin asymmetry, 
$ A^{J_{0}(b_TP_{hT})}_{LL}(b_T)$, defined as
\be
 A^{J_{0}(b_TP_{hT})}_{LL}(b_T) = \frac{ \tilde \sigma^+(b_T) - \tilde \sigma^-(b_T)}{\tilde \sigma^+(b_T) + \tilde \sigma^-(b_T) }=\frac{\tilde \sigma_{LL}(b_T)}{\tilde \sigma_{UU}(b_T)}=\sqrt{1-\varepsilon^2} \frac{\sum_{q}\tilde g^{q}_1(x,z^2b_T^2) \tilde D^{q}_{1}(z,b_T^2)}{\sum_{q}\tilde f^{q}_1(x,z^2b_T^2) \tilde D^{q}_{1}(z,b_T^2)},   
 \label{tildall}
\ee 
 using measured double spin asymmetries as functions of $P_{hT}$~\cite{Avakian:2010ae},  for fixed $x$, $y$, and $z$ bins.
Here $b_T$ is the Fourier conjugate of $P_{hT}$. The Fourier transforms of the helicity-dependent cross sections, $\sigma^{\pm}( b_T)$, can be extracted by integration (analytic models) or summation (for data and MC) over the hadronic transverse momentum, weighted 
by a Bessel function $J_{0}$,
\be
 \tilde \sigma^{\pm}( b_T) \simeq S^{\pm}=\sum_{i=1}^{N^{\pm}} J_{0}(b_T P_{hT,i}) \,  .
 \label{spm}
\ee


\begin{figure}[h]
\begin{minipage}{18pc}
\includegraphics[width=18pc]{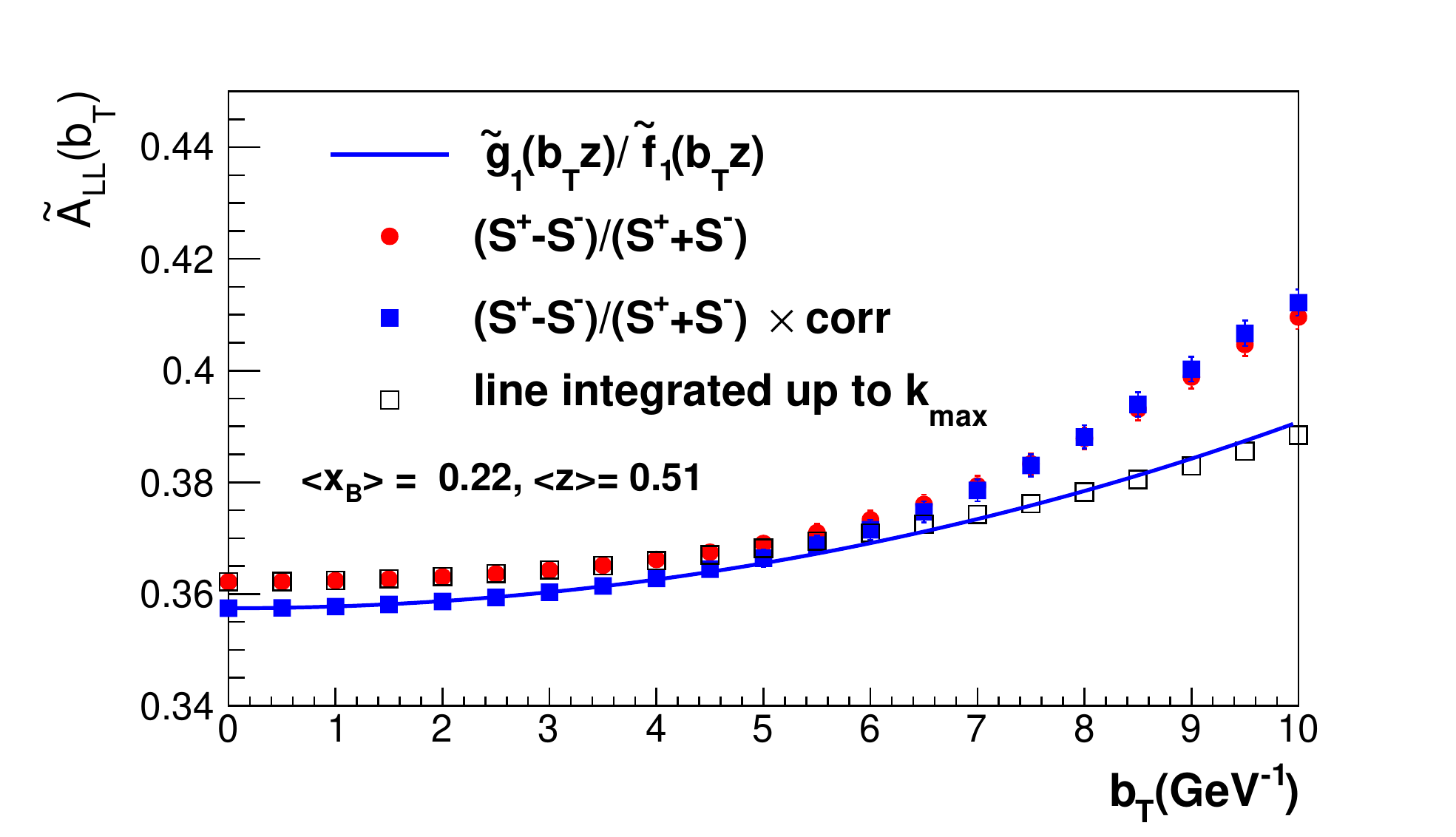}
\caption{\scriptsize  \label{BWPhTcorr} (Color online) Bessel-weighted asymmetry vs $b_T$ with and without the correction, together with analytical and numerical comparison from the MC. See the text for more details. }
\end{minipage}\hspace{2pc}%
\begin{minipage}{18pc}
\includegraphics[width=18pc]{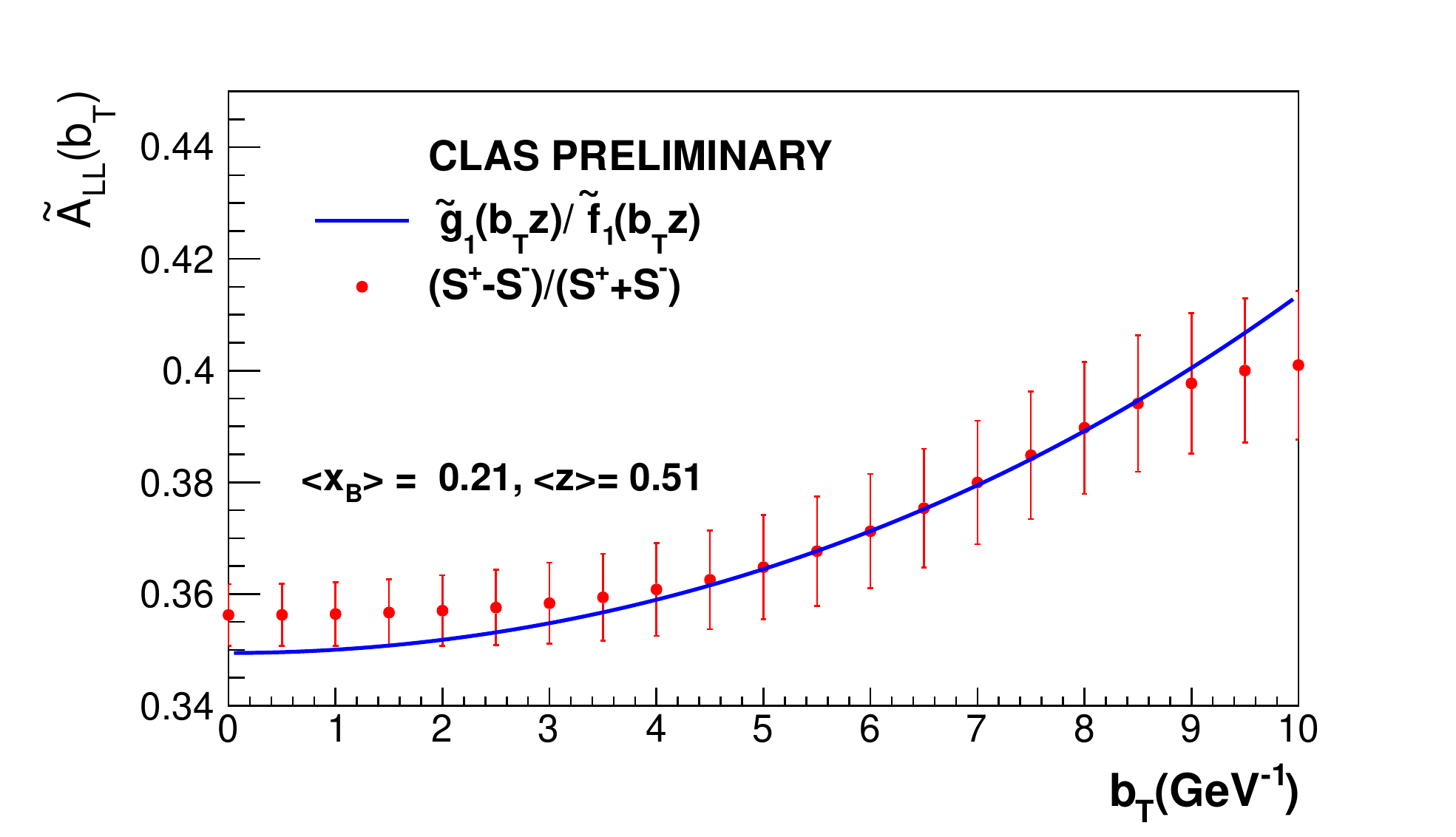}
\caption{\scriptsize \label{BWdata} (Color online) Bessel-weighted asymmetry vs $b_T$ from the electroproduction of neutral pions. Blue line is the same as in the Fig.~2. See the text for more details.}
\end{minipage} 
\end{figure}

The Bessel-weighted asymmetry obtained from the simulated events is shown in Fig.~\ref{BWPhTcorr} as
function of $b_T$ with 
filled (red) circles, while the analytic
expression $\frac{\tilde g_1(x,b_T)}{\tilde f_1(x,b_T)}$ using $\<\kt^2\>_{g_1}$ and $\<\kt^2\>_{f_1}$ 
from the fits to $\kt^2$ distributions from the same MC sample 
is depicted by the (blue) full line.
For values $b_T < 6~{\rm GeV}^{-1}$, which corresponds to about $1~fm$,  the  Bessel-weighted asymmetries could 
be extracted with an accuracy of 2.5\%, although with a systematic shift. 
This clear systematic shift between the extracted and calculated asymmetries
is due to the  kinematic restrictions introduced by energy and momentum conservation, as well as binning effects, which deform the Gaussian shapes 
of the $\kt$ and $\pt$ distributions. 
In experiments, there is always a cutoff at high $P_{hT}$ due to acceptance and the small cross section,  
as well as a cutoff at small $P_{hT}$ where the azimuthal angles are not well defined due to the
experimental resolution. 
These restrictions in $P_{hT}$ directly affect the extracted $\kt$ and $\pt$ distributions and yield the
mentioned distortion of Gaussian shapes which result in the systematic shift. 
Obviously, this shift depends on experimentally introduced restrictions for the accessible $P_{hT}$ range. 

We discussed two approaches in \cite{Aghasyan:2013qqa,Aghasyan:2013kz} which take these conditions into account, one corrects the data (using asummptions) the other applies limits to the integration range
for the intrinsic transverse parton momenta when calculating the asymmetry. 
The model-dependent correction of MC is shown in Fig.~\ref{BWPhTcorr} by the (blue) filled squares. 
The extracted asymmetry now matches the theoretical curve for values $b_T < 6~{\rm GeV}^{-1}$.
Alternatively, 
limited numerical integration over intrinsic transverse momenta in the calculation of the 
asymmetry, where the integration limits correspond to the accessible experimental $P_{hT}$ range, 
yields a calculated asymmetry that describes correctly the experimental situation without introducing 
a model dependence. 
This is shown in Fig.~\ref{BWPhTcorr} by the (black) open squares.

In Fig.~\ref{BWdata} we present a preliminary measurement from the E05113  CLAS dataset of the Bessel-weigthed asymmetry vs $b_T$ for neutral pion electroproduction in SIDIS of 5.8 ${\rm GeV}$ polarized electrons from a longitudinally polarized ammonia target using the CEBAF Large Acceptance Spectrometer (CLAS) at the Thomas Jefferson National Accelerator Facility. The theoretical curve in Fig.~\ref{BWdata} is the same as in the Fig.~\ref{BWPhTcorr}. Deep-inelastic scattering events were selected by requiring $Q^2>1 {\rm GeV}^2$ and $W^2>$ $4 {\rm GeV}^2$. where $W$ is the invariant mass of the hadronic final state. Events with missing-mass values for the $e\pi^0$ system that are smaller than $1.5$ ${\rm GeV}$ ($M_x(e\pi^0)<1.5 {\rm GeV}$) were discarded to exclude contributions from exclusive processes. Error bars are only statistical. An additional $5\%$ scaling uncertainty due to the beam and target polarization measurements should be added. Another source of systematic uncertainty is the dilution factor which is estimated to be within a few percent for each $b_T$ point. The measured Bessel-weigthed asymmetry vs $b_T$ for neutral pions is consistent with the MC points, which is an indication that the ratio of relative widths of  $g_{1L}(x,k_{\perp})$ and $f_{1}(x,k_{\perp})$ in the MC are realistic.
 
 We would like to acknowledge the outstanding efforts of the staff of the Accelerator and the Physics Divisions at JLab that made this experiment possible. This work was supported in part by the National Science Foundation, the Italian Istituto Nazionale di Fisica Nucleare, the Research Infrastructure Integrating Activity
Study of Strongly Interacting Matter (acronym HadronPhysic3, Grant Agreement
n. 283286) under the Seventh Framework Programme of the European Community. The Jefferson Science Associates (JSA) operates the Thomas Jefferson National Accelerator Facility for the United States Department of Energy under contract DE- AC05-06OR23177.

\bibliographystyle{utcaps.bst}

\bibliography{alu}


\end{document}